\documentclass[twocolumn,english,nolinenumbers,aps,pra,twocolumn,unsortedaddress,showpacs]{revtex4-1}
\usepackage[T1]{fontenc}
\usepackage[latin9]{inputenc}
\usepackage{color}
\usepackage{babel}
\usepackage{amsmath}
\usepackage{amssymb}
\usepackage{graphicx}
\usepackage{esint}
\usepackage[unicode=true,pdfusetitle,
 bookmarks=true,bookmarksnumbered=false,bookmarksopen=false,
 breaklinks=false,pdfborder={0 0 0},backref=false,colorlinks=true]
 {hyperref}
\hypersetup{
 citecolor=blue}

\makeatletter

\usepackage{babel}

\graphicspath{{./}{./Figures/}} 

\makeatother

\begin{document}

\title{Electrical Addressing and Temporal Tweezing of Localized Pulses \\
in Passively Mode-Locked Semiconductor Lasers}

\author{P. Camelin}

\affiliation{Université Côte d'Azur, CNRS, Institut Non-Linéaire de Nice, 06560
Valbonne, France}

\author{J. Javaloyes}

\affiliation{Departament de Física, Universitat de les Illes Baleares, C/Valldemossa
km 7.5, 07122 Mallorca, Spain}

\author{M. Marconi}

\affiliation{Université Côte d'Azur, CNRS, Institut Non-Linéaire de Nice, 06560
Valbonne, France}

\altaffiliation{Centre de Nanosciences et de Nanotechnologies, CNRS, Univ. Paris-Sud, Université Paris-Saclay, C2N-Orsay, 91405 Orsay cedex, France}

\author{M. Giudici}

\affiliation{Université Côte d'Azur, CNRS, Institut Non-Linéaire de Nice, 06560
Valbonne, France}
\begin{abstract}
We show that the pumping current is a convenient parameter for manipulating
the temporal Localized Structures (LSs), also called localized pulses,
found in passively mode-locked Vertical-Cavity Surface-Emitting Lasers.
While short electrical pulses can be used for writing and erasing
individual LSs, we demonstrate that a current modulation introduces
a temporally evolving parameter landscape allowing to control the
position and the dynamics of LSs. We show that the localized pulses
drifting speed in this landscape depends almost exclusively on the
local parameter value instead of depending on the landscape gradient,
as shown in quasi-instantaneous media. This experimental observation
is theoretically explained by the causal response time of the semiconductor
carriers that occurs on an finite timescale and breaks the parity
invariance along the cavity, thus leading to a new paradigm for temporal
tweezing of localized pulses. Different modulation waveforms are applied
for describing exhaustively this paradigm. Starting from a generic
model of passive mode-locking based upon delay differential equations,
we deduce the effective equations of motion for these LSs in a time-dependent
current landscape.

\pacs{42.65.Sf, 42.65.Tg, 47.20.Ky, 89.75.Kd}
\end{abstract}
\maketitle

\section{introduction}

Localized structures (LSs) in optical resonators have attracted much
interest in the last twenty years. While LS are ubiquitous in nature
\cite{WKR-PRL-84,MFS-PRA-87,NAD-PSS-92,UMS-NAT-96,AP-PLA-01} and
their investigation conveys an intrinsic fundamental appeal, \emph{optical}
LSs are very attractive also for applications. Because they can be
individually addressed and manipulated, LSs can be used as elementary
bits of information for all-optical information processing \cite{RK-OS-88,TML-PRL-94,FS-PRL-96,1172836}.
Localized structures appear in nonlinear systems having a large aspect-ratio
and they rely on the coexistence between a pattern (pulsed) state
with a stable homogeneous (stationary) one, though different scenarios
exist as well. In the weak dissipative limit, LSs can be interpreted
as dissipative solitons \cite{TF-JPF-88,FT-PRL-90}. 

Localized structures have been observed in optical resonators both
in their transverse section and along their longitudinal dimension.
In the former case they appear as localized beams of light (spatial
LSs) and they have been successfully implemented in broad-area semiconductor
Vertical-Cavity Surface-Emitting Lasers (VCSELs) \cite{BTB-NAT-02,GBG-PRL-08,TAF-PRL-08,EGB-APB-10}.
In the latter case they appear as localized temporal light pulses
(temporal LSs). In the time domain the large aspect-ratio condition
requires the resonator round-trip to be much larger than any internal
timescales of the system, a condition which has been implemented,
for example, in Kerr cavities \cite{LCK-NAP-10,HBJ-NAP-14}. 

Another remarkable property of LSs is related to their plasticity.
Because of translational invariance, LSs exhibit a Goldstone mode
\cite{FS-PRL-96,MFH-PRE-02} which is excited by any inhomogeneous
parameter variation, thus inducing their motion. Because the velocity,
instead of the acceleration, is proportional to the parameter variations,
the latter is interpreted as an Aristotelian force which allows for
LS reconfiguration and drift \cite{FGB-APL-06,PBC-APL-08,JEC-NAC-15}. 

Semiconductor gain media, which typically exhibit nanosecond response
timescales (owing to the recombination rate of carriers), are promising
for achieving large temporal aspect-ratio condition required for temporal
localization. Accordingly, temporal LSs have been recently observed
in semiconductor lasers mounted in compound cavities configurations.
Several regimes have been exploited to provide the dynamical ingredients
leading to LS: front stabilization \cite{MGB-PRL-14,JAH-PRL-15},
passive mode-locking \cite{MJB-PRL-14}, excitability \cite{GJT-NC-15,RAF-SR-16},
and polarizations competition \cite{MJB-NAP-15}. Interesting enough,
the dynamics of these systems is affected by the finite response of
the carriers, i.e. their causal response, thus breaking the parity
invariance along the longitudinal direction of the cavity. 

\begin{figure*}[!t]
\includegraphics[bb=0bp 0bp 733bp 535bp,clip,width=1\columnwidth]{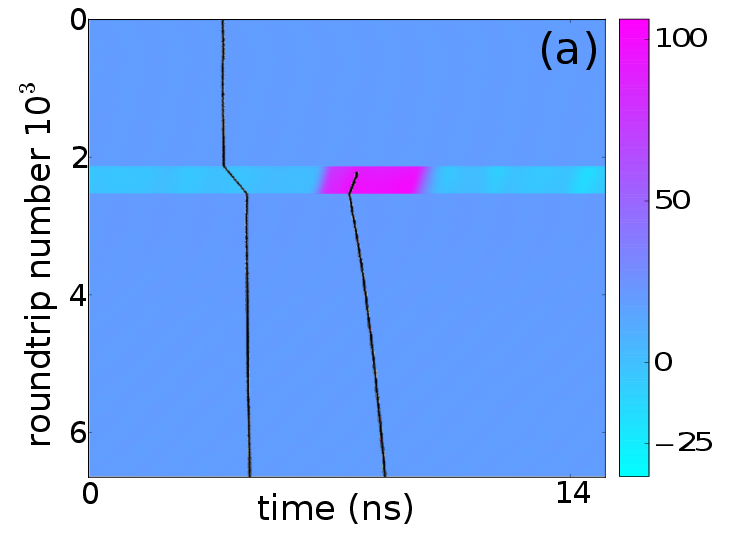}\includegraphics[bb=0bp 0bp 733bp 535bp,clip,width=1\columnwidth]{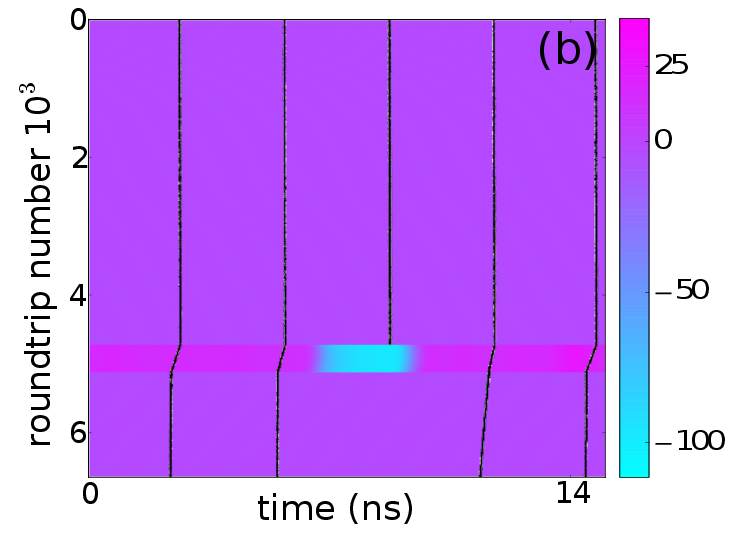}\caption{Spatio-temporal diagrams showing the addressing, i.e. the writing
(a) and the erasing (b) of a LS, by an electrical pulse in the pumping
current. The current values are represented in color scales, while
optical pulses trajectories are represented by black lines. $J_{cw}=211\,$mA.
The amplitude and the width of the writing pulse are $A=130\,$mA
and $2.9\,$ns, respectively. \label{addressing}}
\end{figure*}

In this paper we consider an electrically-biased broad-area VCSEL
coupled to a resonant saturable absorber mirror (RSAM) which leads
to passive mode-locking \cite{MJB-JSTQE-15}. In the limit of cavity
round-trips much longer than the gain recovery time, we have shown
that mode-locked pulses may coexist with the zero intensity background.
In this condition, these pulses become \emph{localized}, i.e. they
are temporal\emph{ }localized structures which can be individually
addressed by a perturbation pulse \cite{MJB-PRL-14}. In the recent
years, temporal LSs have been addressed all-optically by perturbation
of the phase/amplitude of the injected field \cite{LCK-NAP-10,GJT-NC-15}.
On the other hand, in electrically-biased semiconductor lasers, the
pumping current can be easily modulated at several gigahertz rates,
as demonstrated in opto-electronics for converting an electrical bit-stream
into an optical one \cite{Freeman}. Accordingly, the bias current
is a convenient parameter for addressing and manipulating LSs in view
of their applications. We have recently shown that, in a modulated
current landscape, the resulting drifting speed of the localized pulses
is not simply proportional to the gradient of the landscape, as it
occurs in quasi-instantaneous media \cite{JEC-NAC-15}, but it rather
depends on the local value of the parameter itself \cite{JCM-PRL-16}.
This difference, which is ascribed to the lack of parity invariance
described above, leads to a new paradigm for LS manipulation which
will be described in details in this work.

The manuscript is organized as follows : Section II provides a detailed
experimental analysis of the individual writing and erasing of LSs
by using electrical pulses in the pumping current. In section III
we analyze experimentally the motion and the reconfiguration of LSs
in presence of a temporal parameter landscape introduced by modulating
the pumping current with different kinds of waveforms. The theoretical
analysis of these results is described in section IV where the Effective
Equations of Motion (EEM) for multiple LSs in a parameter landscape
are presented and used to explain our experimental results.

\section{Individual addressing of localized pulses}

We consider the setup described in \cite{MJB-PRL-14,MJB-JSTQE-15}
consisting of an electrically biased broad-area ($200\,\mu$m) VCSEL
mounted in an external cavity closed by a resonant saturable absorber
mirror (RSAM). Passive Mode-Locking is achieved when placing the RSAM
surface in the Fourier transform plane of the VCSEL near-field profile.
While this scheme leads to conventional mode-locking pulses for cavity
round-trips shorter than the gain recovery time ($\tau<\tau_{g}$),
we operate the system in the regime of long cavities, i.e. $\tau\gg\tau_{g}$,
and for bias currents below the lasing threshold of the compound system
($J_{th}\sim350\,$mA). In these conditions a large variety of pulsating
emission states coexist, each one characterized by a different number
$N$ of pulses circulating in the cavity or, when having the same
value of $N$, characterized by different pulse arrangements. The
number of pulses $N$ spans from zero up to a maximum $N_{max}$ which
depends on the cavity length. This generalized multistability provides
the dynamical ingredients for the formation of LSs. The state at which
$N=N_{max}$ corresponds to the fully developed temporal pattern which
is, together with the coexisting stable off solution, at the origin
of the LS formation scenario. In analogy to spatial LSs \cite{CRT-C-94,CRT-PRL-00},
the temporal pattern is fully decomposable because any pulse of this
solution can be set on or off by a perturbation pulse \cite{MJB-PRL-14}.
We have fixed our cavity length to 2.25~m, corresponding to a round-trip
time $\tau=15.02\,$ns, leading to $N_{max}=19$.

Writing and erasure of temporal LSs can be implemented by perturbing
the system with an optical pulse injected inside the cavity, as done
in \cite{LCK-NAP-10,GJT-NC-15}. The perturbation pulse must be sufficiently
short in time for addressing a single LS. In this paper we take advantage
of the fast response of semiconductor media to the pumping current
modulation and we apply the addressing perturbation by adding electrical
pulses to the laser bias. This parameter appears the most convenient
in view of opto-electronic applications; conversion of an electrical
bit stream into an optical one has been recently achieved by modulating
laser current at more than 10~GHz \cite{Freeman}. Unfortunately,
because of the electrical characteristics of the laser package we
have used, the modulation of the laser current suffers of stronger
bandwidth limitations with a low frequency cut off at 0.2~MHz and
a high frequency cut off at 500~MHz. Still, this frequency window
is wide enough to allow for a proof-of-principle demonstration of
LSs' addressing operation.

The system is initially prepared in the multistable parameter region
where LSs exist, and the amplitude of the addressing current pulse
is chosen to be sufficiently large to bring the system locally above
threshold, i.e. into the parameter region where only the temporal
pattern solution is stable. If the electrical pulse is sufficiently
short, a single LS will be excited and it will remain after the perturbation
is removed. Similarly, a single localized structure can be erased
by using a negative bias current pulse which brings the system in
the parameter region where only the off solution is stable. In our
case, the electrical pulse has a rectangular shape with a duration
of $2.9\,$ns and it is applied synchronously with the cavity round
trip-time for 390 consecutive round-trips. These two characteristics
of the electrical pulses are imposed by the performances of the pulse
generator used in burst mode. Unfortunately, it was not possible to
obtain shorter burst of pulses, nor shorter pulses.

In order to follow the evolution of LSs inside the cavity we depict
the laser intensity output using the so-called space-time diagrams,
where the time trace is transformed into a two-dimensional representation
with a folding parameter equal to the pulse repetition period $T$.
Accordingly, the round-trip number $n$ becomes a pseudo-time variable
while the pseudo-space variable corresponds to the timing of the pulse
modulo $T$. This representation is similar to the one used to display
the evolution of pulses along optical fi{}bers where fast and slow
timescales are well separated and it has been proposed also for delayed
systems \cite{AGL-PRA-92}, yielding, in some cases, to a direct equivalence
with the Ginzburg-Landau equation \cite{GP-PRL-96}. 

The bias current evolution is represented on the space time diagram
using a color code, while the trajectory of the LS is represented
by a black trace. In Fig.~\ref{addressing}a) we illustrate a writing
operation using a current pulse applied between round-trip $n_{1}=2136$
and round-trip $n_{2}=2526$. We choose an initial condition where
a single LS is present inside the cavity. The perturbation pulse excites
a second LS which remains after the perturbation is removed. Figure~\ref{addressing}a)
is an example of a writing operation starting from a particular initial
condition. Other initial conditions can be chosen with similar results,
provided that the addressing pulse is separated in time from the preexisting
LSs of at least $1\,$ns. On the other hand, we notice that, after
the electrical pulse is removed, the written LS drifts to the right,
thus disclosing the existence of a repulsive force exerted by the
first LS on the written one. This repulsive interaction is asymmetric
since the leftmost LS pushes the rightmost LS but not vice-versa.
The latency in the writing process and the repulsive interaction are
explained by the gain depletion induced by a LS and will be detailled
in Section IV. The gain recovery process follows an exponential law
with a typical time constant of $\Gamma^{-1}\sim1\,$ns. Accordingly,
the gain depression prevents writing two LSs in a time interval smaller
than $\Gamma^{-1}\sim1\,$ns, while the repulsive force between neighbor
LS becomes negligible after $\sim3\Gamma^{-1}$, i.e.$3\,$ns \cite{JCM-PRL-16}.
These evidences indicate that, even if the intensity profile of a
single temporal LS exhibits a temporal width of approximately 10~ps,
the effective LS width is ultimately fixed by the underlying gain
recovery process.

Incidentally one notes in Fig.~\ref{addressing}a) that the trajectory
of the preexisting LS changes when the electrical pulse is applied.
This is explained by the current profile of the perturbation pulse
which, around the 2.9~ns rectangular peak, adds a negative contribution
to the steady bias current, thus changing the drifting speed of the
preexisting LS. The influence of the pumping current landscape on
the drifting speed of the LSs will be addressed in the next section. 

Finally, an example of an erasing operation is shown in Fig.~\ref{addressing}b),
where the initial condition is composed of five preexisting LSs. A
negative pulse of $2.9\,$ns is sent into the laser current between
the round-trips $n_{1}=4710$ and $n_{2}=5100$, targeting the third
LS from the left. It successfully erases the third LS which remains
off after the perturbation is removed. It is worthwhile noting the
reorganization of the positions of the other LSs after the erasing
operation; the fourth LS starts to drift to the left, because the
repulsive force exerted by the erased LS has disappeared. The asymmetrical
interactions between LSs will be discussed more in details in Section
IV.

\section{Motion of localized pulses}

\begin{figure}
\includegraphics[bb=80bp 0bp 1600bp 889bp,clip,width=1\columnwidth]{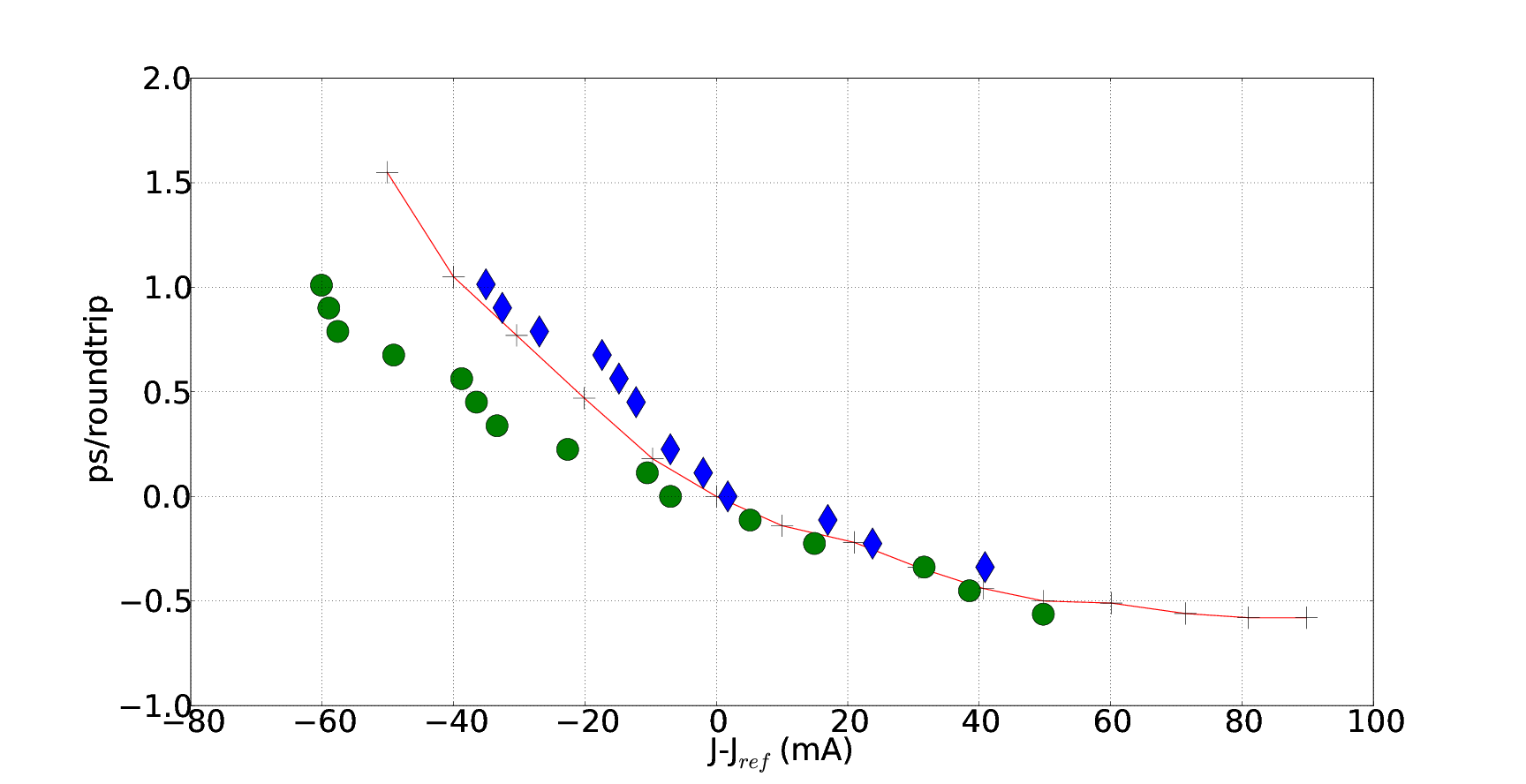}\caption{Drifting speed per round-trip of a LS induced by a bias current variation
around a reference value ($J-J_{ref})$, $J_{ref}=226\,$mA. Bias
current has been changed in different ways: (crosses connected by
red line) by changing its steady value; (circles) by modulating it
sinusoidally, as obtained from Fig.~\ref{sinus}a) and (diamonds)
by modulating it with a triangular waveform, as obtained from Fig.~\ref{triangle}a).}

\label{vsJ}
\end{figure}

\begin{figure*}[t]
\includegraphics[bb=0bp 0bp 2902bp 536bp,clip,width=2\columnwidth]{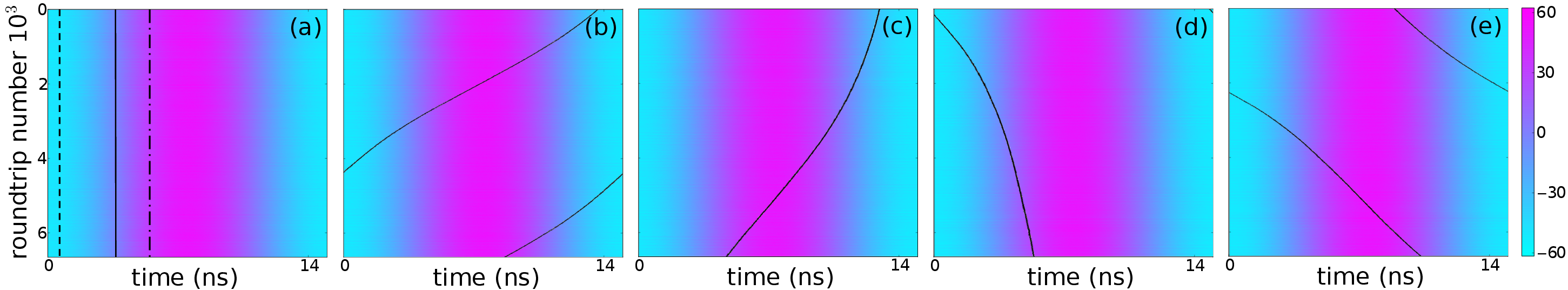}\caption{Spatio-temporal diagram of pulse position evolutions (dark trace)
under a sinusoidal modulation of the current around $J_{ref}=226\,$mA,
at $\nu_{cav}=66614250\,$Hz. The current values are represented in
color scales. a) Three different situations are shown corresponding
to stable equilibrium positions for $\Delta=-0.25\,$kHz (continuous
line), $\Delta=-4.750\,$kHz (dashed line) and at $\Delta=+1.750\,$kHz
(dashed - dotted line). b) $\Delta=-14.25\,$kHz, c) $\Delta=-9.25\,$kHz,
d)$\Delta=3.25\,$kHz and e)$\Delta=10.75\,$kHz. }
\label{sinus}
\end{figure*}

One of the most interesting property of spatial LSs is their plasticity,
i.e. the possibility of manipulating their position by using a parameter
gradient. If LSs are used as bits of information, this property enables
memory reconfiguration and other functionalities as shift-registers
and delay lines \cite{FGB-APL-06,PBC-APL-08,GJT-NC-15,JEC-NAC-15}.
Because of their translational invariance, LSs exhibit a Goldstone
mode \cite{FS-PRL-96,MFH-PRE-02} which is excited by any inhomogeneous
parameter variation, inducing their motion. Since the velocity, instead
of the acceleration, is proportional to the parameter variations,
the latter can be interpreted as an Aristotelian force which allows
for the reconfiguration of the LS ensemble via controlled drifts.
In the case of temporal LSs observed in driven Kerr fiber resonator
\cite{LCK-NAP-10}, a system which is described using Lugiato-Lefever
equation, temporal tweezing of LS has been experimentally demonstrated
and used for bits reconfiguration \cite{JEC-NAC-15}.

At variance with both spatial and temporal LSs described by the Lugiato-Lefever
equation, we have recently shown that the LSs in our system appear
with a finite drifting speed inside the cavity that depends on the
system parameters and, in particular, on the pumping current \cite{JCM-PRL-16}.
This is due to the finite response time of the carriers, which introduces
causality and breaks the parity symmetry along the propagation direction.
In Fig.~\ref{vsJ} (crosses connected by red line) we show experimentally
how the speed of the LSs varies from its value obtained at a reference
pumping current $J_{ref}$ as a function of the steady pumping current
$(J-J_{ref}$). This curve has been obtained by defining a reference
round-trip time $T$ at $J=J_{ref}$. The value of $T$ is determined
as the folding parameter for which the LS trajectory in the space-time
diagram is a straight vertical line or, equivalently, for which the
LS exhibits no residual drift. Clearly, this definition of $T$ is
different from the round-trip time one would calculate from the optical
length of the cavity as it also contains the contribution of the dynamical
nonlinear interactions with the active media. This length cannot be
measured precisely in our experiment and, for this reason, we adopt
an operative definition of $T$, which includes the propagation speed
of a LS inside the cavity at $J=J_{ref}$. Figure~\ref{vsJ} shows
that when the steady current is increased (decreased) with respect
the value $J_{ref}$, the timing of the LS decreases (increases),
which means that the LS acquires a negative (positive) drifting speed
in the space-time diagram obtained using $T$ as the folding time.
It is worthwhile noting that this result cannot be explained in terms
of Joule heating induced by bias current increase. In fact, this would
normally lead to an increase of the optical path length and thus to
an increase of LS timing, exactly the opposite behavior shown in Fig.~\ref{vsJ}.

\begin{figure}[h]
\includegraphics[bb=55bp 0bp 620bp 179bp,clip,width=1\columnwidth]{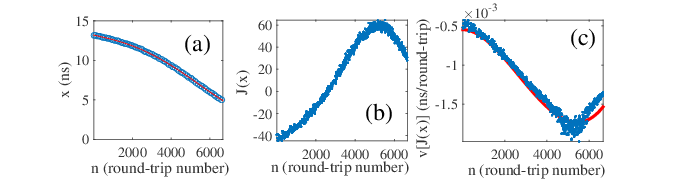}\caption{(a) The blue circles represent the localized pulse position within
the cavity at each round-trip during the trajectory represented in
Fig.~\ref{sinus}c). From these data points a continuous trajectory
is interpolated and represented in red. (b) The values of the current
experienced by the localized pulse at each point of its trajectory.
(c) Drifting speed calculated in two different manners. The blue dots
corresponds to the instantaneous drifting speed calculated as $v_{J}+v_{\Delta}=0$,
and using the values of the bias current experienced by the LS in
panel (b) and where the function $v{}_{J}$ has been obtained by fitting
the curve $v{}_{J}(J-J_{ref})$ shown in Fig.~\ref{vsJ} by circles.
The red curves correspond to the derivative of the interpolated trajectory
in (a). \label{speed}}
\end{figure}

\subsection{Motion induced by a pumping current landscape}

In order to analyze the statics and dynamics of a single localized
pulse in a parameter landscape, we modulate the VCSEL pumping current
within the multi-stable current region where LSs exist. We define
the detuning between the modulation frequency $\nu_{mod}$, and the
cavity resonance frequency $\nu_{cav}=1/T$, as $\Delta=\nu_{mod}$-$\nu_{cav}$.
The value of $T$ is estimated by using the same procedure described
above for a single localized pulse at the steady current value $J=J_{ref}$
around which the modulation is applied.

\begin{figure*}[!t]
\includegraphics[bb=0bp 0bp 2898bp 540bp,clip,width=2\columnwidth]{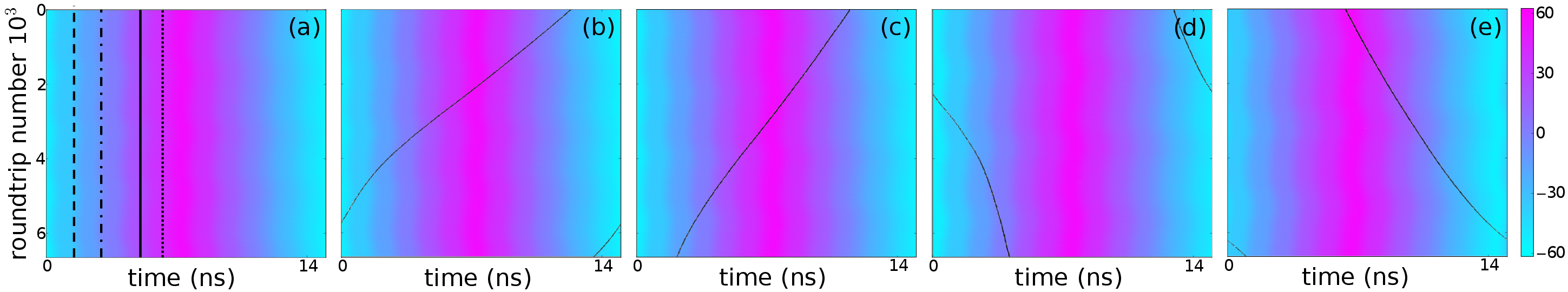}\caption{Spatio-temporal diagram of pulse position evolutions when the current
is modulated by a triangular waveform of amplitude $A=120\,$mA around
$J_{ref}=253\,$mA and at a frequency closed to $\nu_{cav}=66608500\,$Hz.
The current values are represented in color scales. a) Four different
situations are shown corresponding to stable equilibrium positions
for different detuning values: $\Delta=-4\,$kHz (dashed line), $\Delta=-0.5\,$kHz
(dot-dashed line), $\Delta=1\,$kHz (continuous line) and $\Delta=1.5\,$kHz
(dotted line). Drifting regimes with b)$\Delta=-10.5\,$kHz, c)$\Delta=-6\,$kHz,
d) $\Delta=2.5\,$kHz, and e) $\Delta=+6.5\,$kHz. \label{triangle}}
\end{figure*}

\subsubsection{Sinusoidal modulation}

Different modulation signals have been applied, and we start by describing
the results obtained with the sinusoidal one, which can be coupled
into the laser without being distorted up to the 500~MHz cut-off
frequency of our laser module. The modulation amplitude is $A=120\,$mA
around $J_{ref}=226\,$mA. In Fig.~\ref{sinus}a) we show the evolution
of a single pulse for small values of $\Delta$ (-4.75~kHz <$\Delta$<
2.5~kHz) using space-time diagrams. In order to simplify the interpretation
of the LS evolution, we have built these diagrams in the reference
frame of the modulation signal. Accordingly, the current modulation
pattern is always stationary in the space-time diagrams while the
LS position may evolve with the number of round-trips. The current
modulation is represented in Fig.~\ref{sinus}a) using a color scale
and it is superimposed to the LS trajectory in order to make evident
how the localized pulse behaves onto this current landscape. Figure~\ref{sinus}a)
shows that, when $\Delta$ is within the range specified above, the
pulse position is stationary with respect to the modulation signal
and that the stationary position is a function of $\Delta$. 

When $\Delta=0$ the localized peak exhibits a fixed position with
respect to the modulation signal. This position is located on the
positive slope at the point where the modulation signal is equal to
zero. It is important to note that, in this stable position for the
LS, the derivative of the modulation signal is different from zero.
This result points out the difference of our LSs with respect those
described by the Lugiato-Lefever equation, which are expected to have
a zero speed only where the parameter derivative (or its gradient
for spatial systems) is equal to zero \cite{JEC-NAC-15}.

If $\Delta\neq0$, in the reference frame of the modulation signal,
the LS experiences a time slip $\delta t$ at every round trip given
by the difference between the modulation period and the cavity round-trip.
In the space-time representation, this time shift at every round-trip
can be interpreted as a detuning-induced drifting speed $v{}_{\Delta}$
given by $v_{\Delta}$=$\frac{\Delta}{\nu_{mod}\nu_{cav}}$ in the
limit of $\Delta\ll\nu_{cav}$. Figure~\ref{sinus}a), shows that,
if $\mid\Delta\mid$ is small enough (-4.75~kHz <$\Delta$< 2.5~kHz),
the LS remains in a stationary position with respect to the modulation
signal. This equilibrium position gets closer to the modulation peak
(bottom) for increasing positive (negative) value of $\Delta$. Moreover,
all the stable equilibrium positions found varying $\Delta$ are located
on the positive slope of the modulation signal. 

These stationary positions for the LSs in presence of a finite detuning
indicate that $v_{\Delta}$ must be compensated by an opposite drifting
speed $v_{J}$ induced by the current variation around $J_{ref}$,
such that $v_{J}+v_{\Delta}=0$. Accordingly, from the stationary
positions expressed in terms of the values of the current, it is possible
to infer the value of $v_{J}$ as a function of $\left(J-J_{ref}\right)$.
These values are represented in Fig.~\ref{vsJ} using circles. The
curve $v_{J}(J-J_{ref})$, obtained by sinusoidally modulating the
bias current, is rather close to the same curve obtained by changing
the stationary value of J (crosses connected by red line). When considering
that thermal effects induced by current variation become less and
less relevant when the current is modulated at high rates, the proximity
of the two curves is a further confirmation that thermal effects do
not play a relevant role in determining the drifting speed of LS. 

Finally, the fact that $\frac{dv{}_{J}}{dJ}<0$, explains also why
the equilibrium points, i.e. those points satisfying the condition
$v{}_{J}+v{}_{\Delta}=0$, are stable only when located on the positive
slope of the modulation while they are unstable when located on the
the negative slope. In the former case, any perturbation of the LS
position around the fixed points implies a change of the current value
experienced by the localized pulse which generates a restoring force,
thus bringing back the pulse towards the fixed point. In the latter
case, any perturbation implies a change of the current value experienced
by the localized pulse which generates a repelling force, thus pushing
the pulse farther away from the equilibrium point. 

For values of $\Delta$ outside the above specified interval, $v{}_{\Delta}$
cannot be balanced by $v{}_{J}$ at any current values spanned by
the modulation and the LS unlocks and starts to drift in the space-time
diagram, as shown in Fig.~\ref{sinus}b-d). The relative speed is
positive (negative) for positive (negative) value of $\Delta$ and
it depends on the localized pulse position within the cavity. In the
limit where $\Delta$ is very large, $v{}_{J}$ is small compared
with $v{}_{\Delta}$ and the drifting speed is almost constant. For
values of $\Delta$ around $10\,$kHz, $v{}_{J}$ is comparable to
$v{}_{\Delta}$ and the LS accelerate and decelerate as a function
of its position inside the cavity, as shown in Fig.~\ref{sinus}b-e).
By extracting from Fig.~\ref{sinus}c) the time law of the localized
pulse (shown in Fig.~\ref{speed}a), we calculate the corresponding
instantaneous velocity in the space-time diagram (in ps per round-trip)
which is represented in Fig.~\ref{speed}c), red curve. From Fig.~\ref{sinus}c)
it is also possible to extract the values of the current $(J-J_{ref})$
experienced by the localized pulse during its trajectory (Fig.~\ref{speed}b).
The curve in Fig.~\ref{vsJ} obtained for sinusoidal modulations
(circles) can be polynomially fitted and used, together with Fig.~\ref{speed}b),
to calculate the value of $v{}_{J}$ at each round-trip. Finally,
by adding $v{}_{J}$ and $v{}_{\Delta}$, calculated from the value
of $\Delta$, one obtains the blue curve in Fig.~\ref{speed}c),
which gives the instantaneous values of LS drifting speed. Within
the experimental uncertainty, the two curves in Fig.~\ref{speed}c)
coincide, thus evidencing that the instantaneous velocity of the localized
pulse depends exclusively on the value of the modulation signal instead
of its time derivative.

\subsubsection{Triangular modulation}

A triangular waveform has been also used for modulating the pumping
current. This waveform is interesting because its time derivative
features a constant positive value on the ascending triangle slope
and the same, but negative, value on the descending triangle slope.
Time derivative is zero only at the top and the bottom of the modulation.
Accordingly, if the drifting speed was depending on the modulation
signal time derivative, the equilibrium position should be located,
for $\Delta=0$, on the modulation extrema and the drifting speed
should be constant on the slopes of the modulation. Figure~\ref{triangle}a)
shows the stable equilibrium positions for four different values of
the detuning. The stationary positions are located on the positive
slope of the signal and they depend on $\Delta$. This is a further
confirmation of the negligible dependence of $v{}_{J}$ on the modulation
signal derivative, which is a constant on the entire signal positive
slope. Similarly to the case of the sinusoidal modulation, the stationary
positions of the LS with respect to the modulation signal can be used
to infer the value of $v{}_{J}$ induced by a current variation around
the steady value $J_{cw}$. The curve $v{}_{J}(J-J_{ref})$ is again
plotted in Fig.~\ref{vsJ} using diamond markers. The situation is
very similar to the one obtained by varying the continuous value of
the current and the one obtained by modulating sinusoidally the pumping
current. When the detuning is exceedingly large with respect the current
variation, the term $v{}_{J}$ can never compensate for the term $v_{\triangle}$
and the localized pulse drifts inside the cavity with a speed per
round-trip which is a function of the local value of the current,
as shown in Fig.~\ref{triangle}b-e). The variations of the speed
on the signal slopes give a visual indication that $v{}_{J}$ is a
function of current values rather than its derivative.

\begin{figure*}[!t]
\includegraphics[bb=0bp 0bp 2897bp 540bp,clip,width=2\columnwidth]{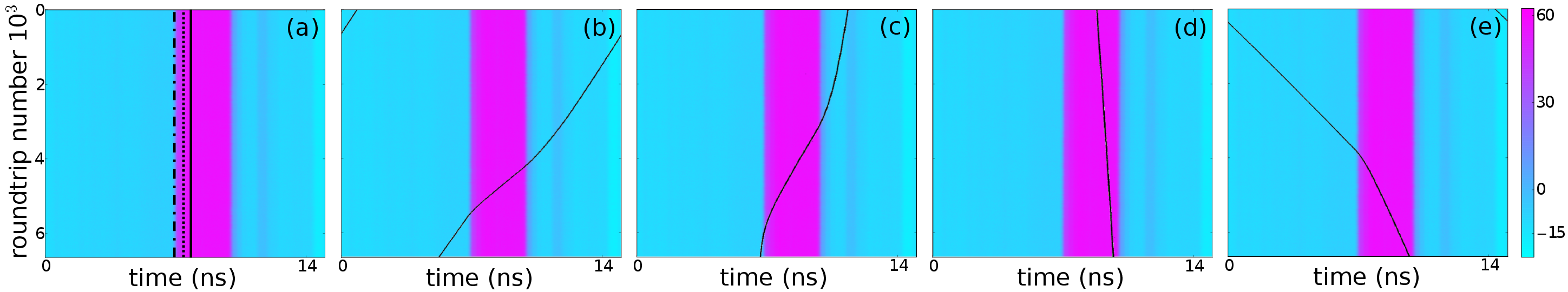}

\caption{Spatio-temporal diagram of pulse position evolutions (dark trace)
when the current is modulated by a rectangular waveform of amplitude
$A=76\,$mA around $J_{cw}=221\,$mA and at a frequency close to the
resonance $\nu_{cav}=66612500\,$Hz. The current values are represented
in color scales. a) Three different situations are shown corresponding
to stable equilibrium positions for different detuning values: $\Delta=-1.3\,$kHz
(dash-dotted line), $\Delta=1\,$kHz (dotted line) and $\Delta=2.5\,$kHz
(continuous line). Out of the locking range, drifting trajectories
are depicted for detuning values of b) $\Delta=-7.5\,$kHz, c) $\Delta=-1.9\,$kHz,
d) $\Delta=3.8\,$kHz and e)$\Delta=7\,$kHz. \label{rect}}
\end{figure*}

\subsubsection{Rectangular modulation}

The experimental evidences reported so far indicate that a rectangular
current modulation, which introduces a bi-valued current landscape,
should enable robust temporal tweezing of localized pulses. Because
the current is widely spanned in the vicinity of the rectangle waveform
edges, the rising-edge of the signal should be a narrow anchoring
region for localized pulses for a wide detuning range. We select a
rectangular waveform almost resonant with the cavity free spectral
range having a $20\,\%$ duty cycle, which means that current is on
the high level for $3\,$ns and on the low level for $12\,$ns. The
rise time of the rising edge is 300~ps (10-90~\%) and 800~ps (0-100~\%).
The current profile is not entirely flat after the steep rising edge,
and it continues to slightly increase during 450~ps. In Fig.~\ref{rect}
we show the current profile in color scale and the evolution of the
localized pulse inside the cavity round-trip after round-trip. In
the static case (Fig.~\ref{rect}a), different trajectories of the
localized pulse are shown for different values of $\Delta$. The deviation
of the current profile from a perfect rectangular shape is not strong
enough to be evident on this figure. On the other hand, this deformation
explains why the stable fixed positions of the localized pulses for
different value of $\Delta$ are not aligned with the rising-edge
but they are rather spread on the top of the square modulation. Despite
this undesired effect, it is reasonable to claim that the region around
the rising-front is an anchoring region for LSs for a wide detuning
range (-1.1~kHz <$\Delta$<2.5~kHz). We notice also that, for the
same reasons explained for the other waveforms, the falling edge of
the pulse only contains unstable equilibrium points. When the detuning
is exceeding the locking range, the localized pulse drifts inside
the cavity, as shown in Fig.~\ref{rect}b-e). The drifting speed
can be roughly considered bi-valued, as it follows the bi-valued current
profile. When the value of $\Delta$ is close to the locking range,
see Fig.~\ref{rect}c), the speed variation of the localized pulse
as it gets close to the edges of the current profile provides a visual
demonstration of smooth rising and fall processes of the pumping current.

\subsection{Manipulation of Localized Pulses}

\begin{figure}
\includegraphics[bb=0bp 0bp 1273bp 535bp,clip,width=1\columnwidth]{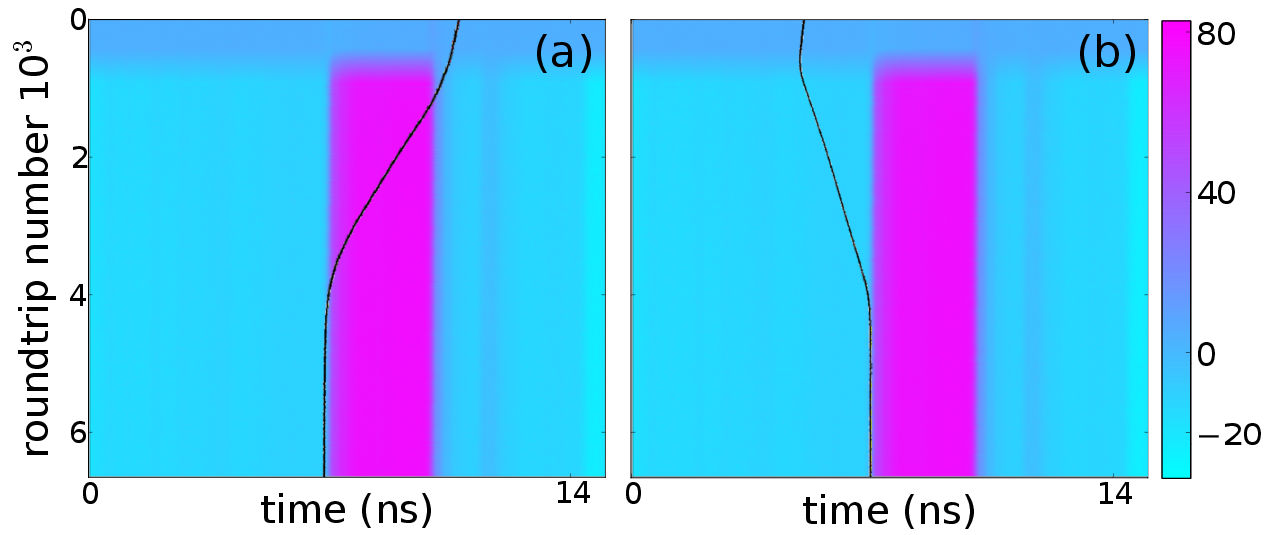}

\caption{Spatio-temporal diagrams showing the evolution of a localized pulse
when an electrical pulse is applied to the pumping current for two
different initial positions. The current values are represented in
color scales. The Current pulse amplitude is $A=100\,$mA, $J_{cw}=202\,$mA,
$\Delta=1.4\,$kHz (left panel) and $\Delta=0\,$kHz (right panel).
\label{buffertrace} }
\end{figure}

The results shown above indicate that current modulation is very effective
for manipulating localized pulses. In Fig.~\ref{buffertrace} we
apply an electrical rectangular waveform to the pumping current and
we study the dynamics induced on the LS. Wherever the LS is located
previously to the application of the current modulation, Fig.~\ref{buffertrace}
shows that it eventually drifts towards the raising edge of the current
pulse where it gets anchored. The trajectory for approaching the rising-edge,
as well as the final position, depend on the value of $\Delta$. In
Fig.~\ref{buffertrace}a), a positive detuning value pushes the LS
towards the rising front and, when it reaches the top of the current
pulse, this speed is increased by the contribution from the high current
level. Eventually the LS sits at the equilibrium point close to the
rising edge of the pulse current. In the right panel, the LS is on
the left of the current pulse. When the pulse is applied, the current
level outside the pulse is slightly decreased with respect to $J_{cw}$
($\sim20\,$\% of the pulse amplitude) and, as a consequence, the
LS experience a drifting speed towards the rising edge of the pulse
current. Eventually it sits in the stable equilibrium point on the
rising edge.

\begin{figure}
\includegraphics[bb=0bp 0bp 750bp 550bp,clip,width=1\columnwidth]{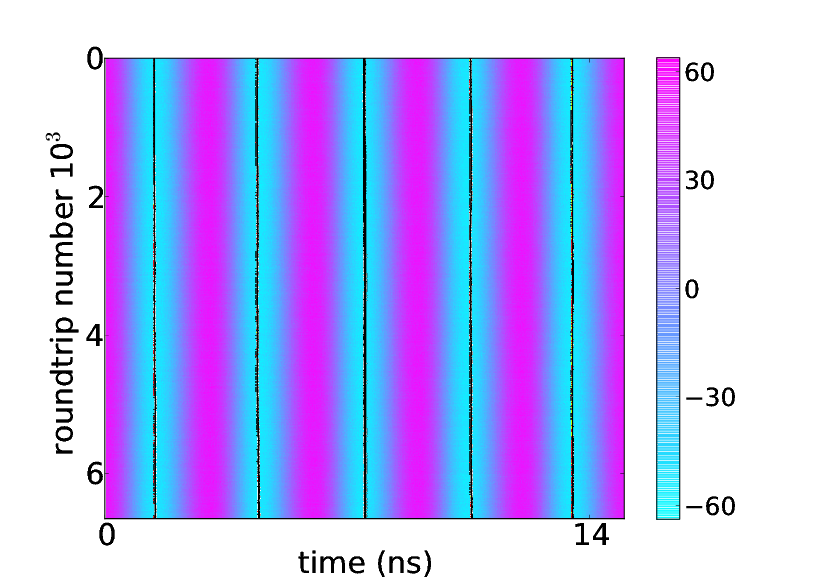}\caption{Spatio-temporal diagrams showing five pulses positions evolution (dark
trace) in presence of a sinusoidal current modulation at $\nu_{mod}=5\nu_{cav}-4.750\,$kHz,
having an amplitude of 120~mA and $J_{cw}=275\,$mA. The current
values are represented in color scales.\label{5f}}
\end{figure}

Localized pulse position manipulation is crucial for applications
to information processing. For example, it is important to organize
the bit flow inside the cavity according to a precise clock. This
can be implemented by introducing a parameter landscape having a period
close to a fraction $N$ of the cavity roundtrip ($\nu_{mod}\approxeq N\nu_{cav}$)
\cite{JEC-NAC-15}. This landscape plays the role of a potential network
capable of trapping $N$ localized pulses at fixed time intervals
inside the cavity. Despite the modulation bandwidth limitation described
in our current set-up, we can still provide a proof-of-principle of
this operation by modulating sinusoidally the pumping current at $N=5$,
i.e. at $\nu_{mod}\approxeq5\nu_{cav}$, which allows pinning the
position of five localized pulses, as shown in Fig.~\ref{5f}. 

\section{Theory}

The existence and dynamics properties of temporal localized structures
in passive mode-locked VCSELs have been theoretically described \cite{MJB-PRL-14,MJB-JSTQE-15,MJC-JSTQE-16}
using the following delay differential equation (DDE) model \cite{VT-PRA-05}:
\begin{eqnarray}
\frac{1}{\gamma}\frac{dA}{dt}+A & = & \sqrt{\kappa}R\left(t-\tau\right)A\left(t-\tau\right),\label{eq:VT1}\\
\frac{dG}{dt} & = & \Gamma\left(G_{0}-G\right)-e^{-Q}\left(e^{G}-1\right)\left|A\right|^{2},\label{eq:VT2}\\
\frac{dQ}{dt} & = & Q_{0}-Q-s\left(1-e^{-Q}\right)\left|A\right|^{2},\label{eq:VT3}
\end{eqnarray}
with $R\left(t\right)=\exp\left[\left(1-i\alpha\right)G\left(t\right)/2-\left(1-i\beta\right)Q\left(t\right)/2\right]$,
$G_{0}$ the pumping rate, $\Gamma=\tau_{g}^{-1}$ the gain recovery
rate, $Q_{0}$ the value of the unsaturated losses which determines
the modulation depth of the SA and $s$ the ratio of the saturation
energy of the gain and of the SA sections. We define $\kappa$ as
the amplitude transmission of the output mirror. In Eqs.~(\ref{eq:VT1}-\ref{eq:VT3})
time has been normalized to the SA recovery time. The frequency filter
(that mimics the gain curve) has a bandwidth of $\gamma$ and the
cold cavity round-trip is given by the inverse of the time delay $\tau$.
The alpha factor of the gain and of the absorber sections are $\alpha$
and $\beta$, respectively. The main usefulness of Eqs.~(\ref{eq:VT1}-\ref{eq:VT3})
is to provide an uniformly accurate description of the whole set of
system solutions (including single- and multi- longitudinal modes
ones), among which the pulsating mode-locked solutions are only a
subset. Accordingly, this model allows to unfold the bifurcation diagram
using DDEBIFTOOL \cite{DDEBT}, thus depicting how the various solution
branches are linked one with the others and how temporal LSs appear
\cite{MJB-PRL-14}.

Elaborating on these results and taking advantage of the long cavities
limit at which we operate this system experimentally, we can reduce
the dimensionality of the model. Such approach was successfully used
in \cite{JCM-PRL-16} and also for the theoretical study of light
bullets \cite{J-PRL-16}. Light bullets, or spatio-temporal localized
structures, have strong link with the current work as they share the
same mechanism for temporal localization. We will show that this approach
allows finding the effective equation of motion for the LS quite directly.
The general theory for the effective equation of motion of LS in DDEs
will be presented elsewhere.

We start by putting the delayed terms in the left hand side of Eq.~\ref{eq:VT1}
and define a smallness parameter as the inverse of the filter bandwidth
setting $\varepsilon=1/\gamma$ to find
\begin{eqnarray}
\varepsilon\frac{dA}{dt}\left(t+\tau\right) & + & A\left(t+\tau\right)=\sqrt{\kappa}R\left(t\right)A\left(t\right).
\end{eqnarray}

Physical intuition dictates that the pulse-width scales as the inverse
of the filter bandwidth and is proportional to $\gamma^{-1}=\varepsilon$.
In a related way, one can foresee that the period of the pulse train
scales as $T_{0}\sim\tau+\gamma^{-1}$, i.e. the period is always
larger than the delay due to causality and the finite response time
of the filtering element. Consequently, the solution is slightly drifting
to the left in a space-time diagram in which horizontally is the local
time, playing the role of space $z$, and vertically (downward) the
round-trip number, playing the role of a slow time $s$. As such,
we assume that the solution is composed of two time scales and write
\begin{eqnarray}
\frac{d}{dt} & \rightarrow & \frac{\partial}{\partial x}+\varepsilon^{2}\frac{\partial}{\partial s}
\end{eqnarray}
 and, following \cite{GP-PRL-96}, we express the delayed term as
\begin{eqnarray}
A\left(t+\tau\right) & = & A\left(x+\varepsilon\upsilon,s+\varepsilon^{2}\tau\right)
\end{eqnarray}
This implies that the solution is allowed to evolve on the slow time
scale but also to drift on the fast one with an unknown drift velocity
$\upsilon$. We find that 
\begin{eqnarray}
A\left(t+\tau\right) & = & A+\varepsilon\upsilon\frac{\partial A}{\partial x}+\frac{\left(\varepsilon\upsilon\right)^{2}}{2}\frac{\partial^{2}A}{\partial x^{2}}\nonumber \\
 & + & \varepsilon^{2}\tau\frac{\partial A}{\partial s}+\mathcal{O}\left(\varepsilon^{3}\right)\\
\varepsilon\frac{dA}{dt}\left(t+\tau\right) & = & \varepsilon\left[\frac{\partial}{\partial x}+\varepsilon^{2}\frac{\partial}{\partial s}\right]A\left(t+\tau\right)\\
 & = & \varepsilon\frac{\partial A}{\partial x}+\varepsilon^{2}\upsilon\frac{\partial^{2}A}{\partial x^{2}}+\mathcal{O}\left(\varepsilon^{3}\right)
\end{eqnarray}
yielding up to $\mathcal{O}\left(\varepsilon^{3}\right)$ 
\begin{eqnarray}
\left(\varepsilon\frac{d}{dt}+1\right)A\left(t+\tau\right) & = & A+\varepsilon\left(\upsilon+1\right)\frac{\partial A}{\partial x}\\
 & + & \left[\frac{\left(\varepsilon\upsilon\right)^{2}}{2}+\varepsilon^{2}\upsilon\right]\frac{\partial^{2}A}{\partial x^{2}}+\varepsilon^{2}\tau\frac{\partial A}{\partial s}\nonumber 
\end{eqnarray}

Since we introduced a drift term in order to obtain steady solutions,
we consistently cancel the first order spatial derivative and set
$\upsilon=-1$. In other words, the solution at the next round-trip,
is $A\left(z-\gamma^{-1}\right)$ that is to say, it is shifted to
the left which precisely corresponds to a period of $T_{0}=\tau+\gamma^{-1}$.

Interestingly, we find that the pulse filtering is represented by
a diffusion term of magnitude $\upsilon^{2}/2$ highlighting that
filtering and period deviation with respect to $\tau$ are intimately
related and eventually boil down to causality and finite response
time of the filtering element. In this approach where we factored
out the drift of the filtering element, the residual drift terms will
only be due to the nonlinear interactions with the gain and the absorber. 

We finally find, redefining the time by the value of the round-trip
as $\sigma=\varepsilon^{-2}s/\tau$ and setting $I=\left|A\right|^{2}$
\begin{eqnarray}
\frac{\partial A}{\partial\sigma}-\frac{1}{2\gamma^{2}}\frac{\partial^{2}A}{\partial z^{2}} & = & \left\{ \sqrt{\kappa}e^{\frac{1-i\alpha}{2}G-\frac{1-i\beta}{2}Q}-1\right\} A,\label{eq:VT1a}\\
\frac{\partial G}{\partial z}+\frac{1}{\tau}\frac{\partial G}{\partial\sigma} & = & \Gamma\left(G_{0}-G\right)-e^{-Q}\left(e^{G}-1\right)I,\label{eq:VT2a}\\
\frac{\partial Q}{\partial z}+\frac{1}{\tau}\frac{\partial Q}{\partial\sigma} & = & Q_{0}-Q-s\left(1-e^{-Q}\right)I,\label{eq:VT3a}
\end{eqnarray}
The equations (\ref{eq:VT1a}-\ref{eq:VT3a}) are subject to the following
periodic boundary condition
\begin{eqnarray}
A\left(z+T_{0},t\right) & = & A\left(z,t\right)
\end{eqnarray}
The Eqs.~(\ref{eq:VT1a}-\ref{eq:VT3a}) can be understood as a generalization
of the Haus master equation to large gain and absorption which is
in itself interesting. Indeed, one of the main advantage of the model
of \cite{VT-PRA-05} is the consideration of large gain and absorption,
a feature that is still preserved by the exponential terms in the
above PDE.

In order to reduce further the complexity of the model presented in
Eqs.~(\ref{eq:VT1a}-\ref{eq:VT3a}), we take the limit of small
gain and absorption setting $G\sim Q\sim0.1$ which is still a sensible
approximation even in the case of semiconductor lasers, so that we
can expand the exponential to first order and we also take the long
delay limit in which $\tau^{-1}\partial_{\sigma}\left(G,Q\right)\ll1$
allowing us to keep only the spatial derivative contribution in the
gain and the absorber to find with $d=\left(2\gamma^{2}\right)^{-1}$
\begin{eqnarray}
\negthickspace\negthickspace\negthickspace\negthickspace\negthickspace\negthickspace\frac{\partial A}{\partial t} & \negthickspace=\negthickspace & \left[\sqrt{\kappa}\left(1+\frac{1-i\alpha}{2}G-\frac{1-i\beta}{2}Q\right)-1+d\frac{\partial^{2}}{\partial z^{2}}\right]\negthickspace A,\label{eq:HF1}\\
\negthickspace\negthickspace\negthickspace\negthickspace\negthickspace\negthickspace\frac{\partial G}{\partial z} & \negthickspace=\negthickspace & \Gamma G_{0}-G\left(\Gamma+\left|A\right|^{2}\right),\frac{\partial Q}{\partial z}\negthickspace=\negthickspace Q_{0}-Q\left(1+s\left|A\right|^{2}\right),\label{eq:HF2}
\end{eqnarray}
which was the model directly used in \cite{JCM-PRL-16}.

\subsection{Single LS solution branch }

\begin{figure}
\begin{centering}
\includegraphics[bb=0bp 0bp 520bp 228bp,clip,width=1\columnwidth]{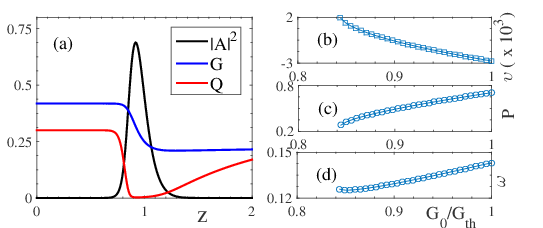}
\par\end{centering}

\centering{}\caption{(a) Localized pulse profile in terms of the field intensity, the carrier
and the absorption with $G=0.99G_{th}$. Panels (b), (c) and (d) depict
the drift velocity $\upsilon$, the pulse integrated energy $P$ and
the residual frequency $\omega$ along the stable solution branch.
Parameters are $\left(\gamma,\kappa,\alpha,\beta,\Gamma,Q_{0},s\right)=\left(40,0.8,1,0.5,0.04,0.3,30\right)$.
\label{Fig:pulsedetail}}
\end{figure}

We search for solutions of Eqs.~(\ref{eq:HF1}-\ref{eq:HF2}) with
a possible drift $\upsilon$ and a carrier frequency $\omega$ as
\begin{eqnarray}
A\left(z,\sigma\right) & = & A_{0}\left(z-\upsilon\sigma\right)e^{-i\omega\sigma}
\end{eqnarray}
which adds in the right-hand side of Eq.~(\ref{eq:HF1}) a term $\left(\upsilon\partial_{z}+i\omega_{0}\right)A_{0}$.
Here, the influence of the drift $\upsilon$ results from the nonlinear
interaction with the active media that have non-instantaneous responses.
Although the full bifurcation diagram of Eq.~(\ref{eq:HF1}) could
be obtained with software as pde2path \cite{pde2path}, as we restrict
our analysis to the stable solution branch, a bifurcation diagram
can be obtained by direct numerical integration. We solved the Eqs.~(\ref{eq:HF1}-\ref{eq:HF2})
via a Fourier based semi-implicit split-step method as in \cite{MJB-JSTQE-15}.
A typical profile can be observed in Fig.~\ref{Fig:pulsedetail}.
Here, the value of the solution parameters $\left(\upsilon,\omega\right)$
were determined self-consistently by imposing the stationarity of
$A_{0}$ as well as a proper phase condition. It is interesting to
notice how this method reduces the complexity of the problem as compared
with the original formulation of the problem as a DDE system. Here,
one does not need to consider a full domain of size $T_{0}\sim\tau$
corresponding to the full extent of the time period imposed by the
large time delay. In our case, the effective numerical domain is kept
only a few time larger than the optical pulse, which remains however
much smaller than the actual extent of the LS, that is governed by
the recovery of the slowest variable $G$. Since the field decays
to zero before and after the emission of the LS, it is not affected
by the ``wrong'' periodic boundary conditions imposed by the spectral
algorithm we used.

A typical LS profile as well as its bifurcation diagram are depicted
in Fig.~\ref{Fig:pulsedetail}. A typical asymmetrical profile can
be observed in Fig.~\ref{Fig:pulsedetail}(a) while the bifurcation
diagrams for $\upsilon$, $P$ and $\omega$ are depicted in Fig.~\ref{Fig:pulsedetail}(b-d).
Importantly, the drift velocity (or equivalently the deviation of
the period with respect to the period $T_{0}$) is a decreasing function
of the bias current, as found experimentally. As we will show later,
this implies that the interaction between distant LSs is purely repulsive.
One notices in Fig.~\ref{Fig:pulsedetail}c) that the solution branch
for the energy of the pulse as defined as $P=\int_{-\infty}^{\infty}|A\left(z,t\right)|^{2}dz$
shows the typical square-root behavior consistent with the fact that
the single LS solution arises as a saddle-node bifurcation of limit
cycle, as demonstrated in the bifurcation diagram of Eqs.~(\ref{eq:VT1},\ref{eq:VT3})
discussed in \cite{MJB-PRL-14}.

The fact that the solution branch exhibits a predominant drift in
a particular direction can be ascribed to the lack of spatial reversibility
$z\rightarrow-z$ in our system that is apparent in the carrier equations
Eqs.~(\ref{eq:HF2}) having only first order derivative in space.
It is worthwhile noting that an adiabatic elimination of the carrier
over the fast temporal variable ($z$), leading to $G=G_{0}/\left(1+|A\left(z,t\right)|^{2}/\Gamma\right)$,
and, similarly for the absorption, would cancel such a drifting solution
and reduce Eq.~(\ref{eq:HF1}) to a generalized Ginzburg-Landau Equation
as in \cite{RK-OS-88}. While the latter adiabatic elimination of
$Q$ could represent some realistic experimental conditions, the former
elimination of $G$ is incorrect in a semiconductor medium since the
recovery time of the gain $\tau_{g}$ is much longer than the pulsewidth
$\tau_{p}\sim\gamma^{-1}$.

\subsection{Derivation of the EEM}

We now outline our approach for deriving the EEM. Well separated LSs
interact via the overlap of their tails. In particular, the interaction
between two consecutive LSs is mediated by the left (resp. right)
tail of the rightmost (resp. leftmost) LS. As the LSs considered in
this manuscript are multi-components and multiscales objects, their
interactions are somehow peculiar. The inspection of Fig.~\ref{Fig:pulsedetail}a)
shows that the rising front is very short; it is governed by the timescale
related to the optical pulse, which is of the order of a few picoseconds
and controlled by the bandwidth of the filtering element $\gamma$.
The black line in Fig.~\ref{Fig:pulsedetail}a) corresponds to an
intensity rising as $I\sim\exp\left(\gamma t\right)$. Oppositely,
the right decaying tail is dominated by the gain recovery that scales
as $G\sim\exp\left(-\Gamma t\right)$, which can be appreciated by
the almost horizontal blue line in Fig.~\ref{Fig:pulsedetail}(a)
following the gain depletion. As the difference between $\gamma$
and $\Gamma$ amounts to several order of magnitudes ($\gamma/\Gamma\sim10^{3}$),
the leftmost LS will influence the rightmost one via its tail induced
by the gain recovery, while the reciprocal interaction will be completely
negligible. In this respect, and as compared to the standard cases
in the literature regarding the interaction between cavity solitons
\cite{MBH-PRE-00}, our situation is counter-intuitive as the interactions
violate completely the action-reaction principle. In our case, the
``reaction'' is completely negligible.

\begin{figure*}
\begin{centering}
\includegraphics[bb=110bp 0bp 960bp 230bp,clip,width=2\columnwidth]{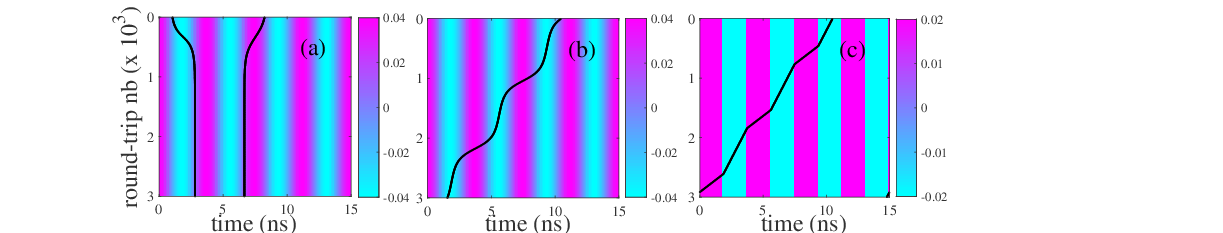}
\par\end{centering}

\centering{}\caption{(a) Resonant harmonic potential: The LS goes to the rising front and
escape of the falling front that corresponds to the stable and the
unstable fixed points, respectively. (b) Strongly detuned harmonic
potential; The LS motion corresponds to unlocked trajectories. Here,
the LS slows down around and accelerate periodically. (c) Strongly
detuned motion beyond the locking range but in a square potential:
Note how the speed is simply bi-valued for the square modulation yielding
a broken line trajectory. The frequency of the external modulation
corresponds to the fourth harmonic $\omega_{ext}=4/T_{0}$ with $T_{0}=15\,$ns.
The amplitude of the harmonic (a-b) and square (b) modulation is $A=0.04$
and $A=0.02$, respectively. The detuning between the natural LS motion
and the external periodic potential are represented by different values
of $\upsilon_{\Delta}$ and reads (a) $\upsilon_{\Delta}=-2.35\times10^{-3}$
(no detuning), (b) $\upsilon_{\Delta}=-1\times10^{-3}$ and (c) $\upsilon_{\Delta}=-1.3\times10^{-3}$.
The current values are represented in color scales.\label{Fig:perio_mouv}}
\end{figure*}

We can use the fact that the interactions will be dominated by the
slowest variable $G$ to build the EEM simply, without the need to
solve the adjoint problem associated with Eqs.~(\ref{eq:HF1},\ref{eq:HF2})
as detailed for instance in \cite{VFK-JOB-99,MBH-PRE-00}. Close to
a saddle-node bifurcation, an excellent approximation is found for
the drift velocity and the pulse energy with a square root ansatz
and we denote 
\begin{eqnarray}
\upsilon & =f\left(G_{0}\right)= & \upsilon_{0}+\Delta\upsilon\sqrt{G_{0}-G_{sn}}\label{eq:upsilon}\\
P & =h\left(G_{0}\right)= & P_{0}+\Delta P\sqrt{G_{0}-G_{sn}}\label{eq:P}
\end{eqnarray}
with the coefficients $\left(\upsilon_{0},\Delta\upsilon,P_{0},\Delta P\right)=\left(2\times10^{-3},-10^{-2},0.28,0.7\right)$
and $G_{sn}=0.845G_{th}$ given by the best fit of the bifurcation
diagram of a single LS in Fig.~\ref{Fig:pulsedetail}. From these
results we can now obtain the EEMs for an ensemble of interacting
LS in a periodic potential directly. We allow for the general case
of a periodic potential whose period is slightly different than the
natural resonance of the system. By going in the reference frame of
the external periodic potential, the equation for the relative drift
velocity Eq.~(\ref{eq:upsilon}) simply becomes 
\begin{eqnarray}
\tilde{\upsilon}+\upsilon_{m} & = & f\left(G_{0}\right)
\end{eqnarray}
with $\upsilon_{m}$ the velocity of the drifting potential and $\tilde{\upsilon}$
the residual speed due to the interactions of the pulse with the gain.
In this reference frame, the external potential depends only on the
fast time $\left(z\right)$ and not anymore on the slow time $\left(\sigma\right)$.
Our main hypothesis consists in saying that, with a spatially dependent
gain, the solution are now depending on the local value of the gain
at the leading edge of the pulse $G^{\left(i\right)}$, i.e. we replace
in Eqs.~(\ref{eq:upsilon},\ref{eq:P}) $G_{0}\rightarrow G^{\left(i\right)}$.
By allowing each LS to imprint a gain depletion and considering the
partial gain recovery in-between LS, we will automatically obtain
the EEMs. We recall that the gain at the falling edge of the pulse
is simply 
\begin{eqnarray}
G^{\left(f\right)} & = & G^{\left(i\right)}\exp\left(-P\right).
\end{eqnarray}
In between LS, during the so-called slow stages where the gain recovers,
the evolution of $G$ is governed by 
\begin{eqnarray}
\frac{\partial G}{\partial z} & = & \Gamma\left[G_{0}\left(z\right)-G\right],
\end{eqnarray}

Provided that the variations of the bias current $G_{0}\left(z\right)$
are slower than the typical relaxation time $\tau_{g}$, i.e. $\Gamma\dot{G}_{0}\ll G_{0}$,
the solution of the carrier equation Eq.~(\ref{eq:HF2}) reads
\begin{eqnarray}
G\left(z_{2}\right) & = & G\left(z_{1}\right)e^{-\Gamma\Delta z}+G_{0}\left(z_{2}\right)\left(1-e^{-\Gamma\Delta z}\right)
\end{eqnarray}
where $\Delta z=z_{2}-z_{1}$ corresponds to the ``distance'' between
two LSs and $G\left(z_{1}\right)$ is the initial condition. Hence,
by denoting as $z_{n}$ the position of the $n$-th LS whose residual
velocity is $\tilde{\upsilon}_{n}=dz_{n}/d\sigma$, we find that
\begin{eqnarray}
\frac{dz_{n}}{d\sigma} & = & f\left(G_{n}^{\left(i\right)}\right)-\upsilon_{m}\;,\; P_{n}=h\left(G_{n}^{\left(i\right)}\right)\label{eq:eqeff1}\\
G_{n}^{\left(i\right)} & = & G_{n-1}^{\left(i\right)}\exp\left[-P_{n-1}-\Gamma\left(z_{n}-z_{n-1}\right)\right]\nonumber \\
 & + & G_{0}\left(z\right)\left(1-\exp\left[-\Gamma\left(z_{n}-z_{n-1}\right)\right]\right)\label{eq:eqeff2}
\end{eqnarray}
where we replaced the initial condition $G_{n-1}^{\left(f\right)}$
at the falling edge of the $\left(n-1\right)$-th LS located in $z_{n-1}$
by $G_{n-1}^{\left(f\right)}=G_{n-1}^{\left(i\right)}\exp\left(-P_{n-1}\right)$.
For $N-$LS with $n\in\left[1,\dots,N\right]$, the periodic boundary
conditions linking the gain depletion of the rightmost LS to the dynamics
of the leftmost one reads 
\begin{eqnarray}
z_{0} & = & z_{N}-T_{0}.
\end{eqnarray}

Some general considerations regarding the EEMs can already be drawn.
The unidirectional interaction between LS is visible in Eq.~(\ref{eq:eqeff2})
where $G_{n}^{\left(i\right)}$ only depends on $z_{n}$, $z_{n-1}$
but not $z_{n+1}$. Also, a great advantage of the EEMs is that the
problem does not become more stiff when the value of the delay increases;
While the cost of a direct integration of Eqs.~(\ref{eq:VT1}-\ref{eq:VT3})
scales linearly with the value of the time delay, the one of the EEMs
remains constant. Secondly, the dynamics of a large ensemble of interacting
LS, possibly in a periodic potential and in the presence of noise,
can be studied via a molecular dynamics approach where each LS is
represented by a point like particle in interactions with its nearest
neighbors. Finally, long simulations are desirable if one wishes to
compare with experimental results, in which the acquisition time amounts
to $10^{4}\sim10^{5}$ round-trips, which can be done with the EEMs
easily. If not otherwise stated, the parameters of the EEMs are $\left(\upsilon_{0},\Delta\upsilon,P_{0},\Delta P\right)=\left(0,-0.01,0.28,0.7\right)$
while $G_{sn}=0.845G_{th}$ and the current value is $G_{0}=0.9G_{th}$.
These parameters were obtained making a best fit of Fig.~\ref{Fig:pulsedetail}b-d).

\subsection{Motion in a periodic potential}

In the case where the potential period is resonant with that of a
single LS, it can be demonstrated from a linear stability analysis
that the stable fixed point corresponds to the zero of the rising
front of the modulation. On the contrary, the falling edge is a saddle
point, due to the fact that $\Delta\upsilon<0$ in Eq.~(\ref{eq:upsilon}).
Figure~\ref{Fig:perio_mouv}a) shows how two initial conditions on
each side of a saddle points over the falling edges, give rise to
two different trajectories. In presence of a detuning between the
modulation and the natural period of the LS motion, the stable and
the unstable equilibrium positions grow closer up to their disappearance
in a saddle-node bifurcation, very much similar to the Adler bifurcation
to unlocked states. A typical unlocked trajectory is depicted in Fig.~\ref{Fig:perio_mouv}b).
Finally, we represent in Fig.~\ref{Fig:perio_mouv}c) the unlocked
motion in a bi-valuated square potential.

\subsection{Interactions between LSs}

We start by considering the case of the self-interaction of a single
LS. The gain depletion generated in the wake of a LS has to be connected
to itself after a single round-trip, we therefore set $z_{n}-z_{n-1}=T_{0}$
and $G_{n}=G_{1}\,\forall n$ which allows us to find solving Eq.~(\ref{eq:eqeff2})
the value of the leading edge gain 
\begin{eqnarray}
G_{1} & = & r_{1}G_{0},
\end{eqnarray}
with the following expression for $r_{1}$ 
\begin{eqnarray}
r_{1} & = & \frac{1-e^{-\Gamma T_{0}}}{1-e^{-\Gamma T_{0}-P}}.\label{eq:Fcrowd}
\end{eqnarray}
The factor $r_{1}<1$ represents the gain reduction induced by the
presence of a single LS as compared to the nominal value $G_{0}$
found only when the cavity is empty. It represents the self-crowding
induced by the LS due to its own gain saturation, if the cavity is
not sufficiently long. This effect of course disappears in the long
delay limit $\Gamma T_{0}\gg1$ where $r_{1}\rightarrow1$.

\begin{figure}
\begin{centering}
\includegraphics[bb=70bp 0bp 770bp 199bp,clip,width=1\columnwidth]{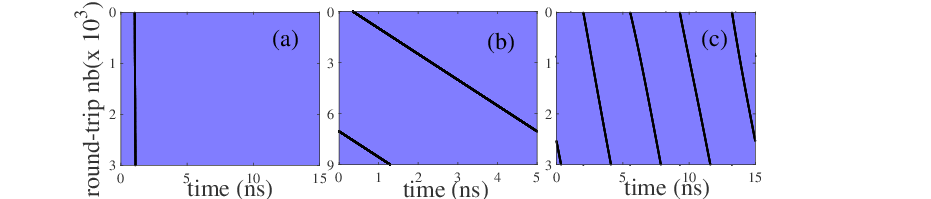}
\par\end{centering}

\centering{}\caption{Effect of self and mutual interactions between LS. Notice how the
drift velocity is affected by the effect of self interactions as the
only difference between panels (a) and (b) is the length of the cavity.
Mutual interactions in a regular crystal structure presented in (c)
also influence the drifting speed with respect to (a). \label{Fig:crowding}}
\end{figure}

A similar result can be found for $N$ equidistant LSs for which we
simply replace $T_{0}\rightarrow T_{0}/N$ in Eq.~(\ref{eq:Fcrowd})
which allows finding $r_{N}$. In particular, one deduces that the
$N-$LSs solution will have a different gain than a single LS solution.
This effect can be linked to the global coupling found for spatial
LSs which was shown to induce a slanting in their bifurcation diagram,
the so-called homoclinic snaking \cite{FCS-PRL-07}. Besides defining
a more limited domain of existence, which was studied in \cite{MJB-PRL-14},
see Fig.~2d, it will also induce a change in their drift velocity.
As the speed $\upsilon$ is a function of $G$, this effect can be
appreciated in Fig.~\ref{Fig:crowding}a,c), where one can see that
the solutions with $N=1$ and $N=4$ have different drifts. In addition,
the difference of gain and therefore of drifting speed as a function
of the cavity length is depicted in Fig.~\ref{Fig:crowding}b) for
the case with $N=1$.

\begin{figure}
\begin{centering}
\includegraphics[bb=0bp 0bp 386bp 252bp,clip,width=1\columnwidth]{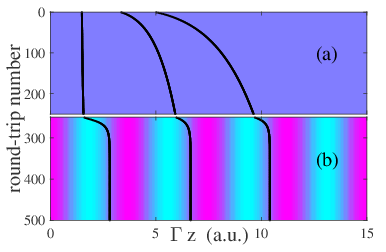}
\par\end{centering}

\centering{}\caption{(a) Asymmetrical interactions between nearby LSs. As soon as an external
modulation is applied (b) the LSs fall at the bottom of the induced
potential. The amplitude of the sinusoidal modulation is $A=0.04$
and the current values are represented in color scales. \label{Fig:asym_int}}
\end{figure}

Finally, we conclude our analysis in Fig.~\ref{Fig:asym_int}a),
where the asymmetrical repulsive interaction between nearby LSs is
made apparent. Here the first LS affects the second one but not the
other way around. The source of the asymmetry is contained in Eq.~(\ref{Fig:asym_int})
where each LS with position $z_{n}$ is only coupled to its previous
neighbor $z_{n-1}$. This theoretical prediction can be compared with
experimentally obtained trajectories of multiple localized pulses
(Fig.~\ref{addressing}) where the asymmetric character of the interaction
is evident. However, these interactions are weak and occurs on very
long time scales. As soon as a external modulation is applied as in
Fig.~\ref{Fig:asym_int}b), the LSs dynamics becomes dominated by
the interaction potential.

\section{Conclusions}

In this manuscript we have shown that temporal LSs in a passively
mode-locked semiconductor laser system can be addressed by using current
electrical pulses. Despite the limited electrical bandwidth of our
laser module, which prevent us from addressing more than one localized
pulse per round-trip, it is important to underline that this method
can lead to a robust and a convenient LS addressing scheme at GHz
rates. These localized pulses exist for a wide range (typically more
than 100~mA) of the VCSEL pumping current $J$ and, similarly to
spatial LS, they can be used as bits for information processing. A
bit stream can be written inside the cavity at an addressing rate
fixed by the temporal width of the LS. More precisely, even if in
terms of the output intensity the LS obtained exhibit a temporal width
of approximately 10~ps, their width is ultimately fixed by the underlying
gain recovery process which has an effective exponential time constant
of a few nanoseconds. As the plasticity of the LSs spatial configuration
renders their individual control possible, the laser bias current
has been used for introducing a parameter landscape and controlling
the position of the localized pulses inside the cavity. In this respect,
we have disclosed a new paradigm for LS manipulation where the induced
drifting speed does not depend upon the gradient of the parameter
landscape but, instead, on its local value, which we traced back to
the strong asymmetry of the LS temporal profile whose rise and fall
time are typically in the picosecond and the nanosecond range, respectively.
We have provided a proof of principle for localized pulse position
reconfiguration using electrical square modulation and we have shown
that a weak amplitude current modulation can be used to precisely
clock the position of the flow of localized pulses inside the cavity.
Our experimental findings where supported by a theoretical analysis
based upon a generic delay differential equation model. However as
the LS dynamics evolves significantly only over a very large number
of round-trips, that are in addition very long as compared to the
usual situations, we simplified our DDE model into an equivalent Haus
master equation that we were able to solve efficiently by exploiting
the long delay limit. This allowed to reduce the numerical domain
to a few times the extend of the optical component of the LS, and
neglect the long tail of the gain recovery. This tail can be obtained
analytically and its asymptotic value used to set the proper boundary
conditions. From the evolution of the LS drift velocity as a function
of the gain, we were able to write the effective equation of motions
for each LS as it they were rigid pseudo-particles governed only by
weak nearest neighbor interactions. These analytical results enabled
us to confirm the most important experimental findings, namely the
presence of strongly asymmetrical interactions and the existence of
equilibrium points in the vicinity of the rising front of the external
modulation.
\begin{acknowledgments}
J.J. acknowledges financial support from the Ramón y Cajal fellowship
and project COMBINA (TEC2015-65212-C3-3-P). The INLN Group acknowledges
funding of Région PACA with the Projet Volet Général 2011 GEDEPULSE
ANR project OPTIROC. M. Giudici thanks the University of Balearic
Islands for a one month visiting position. P. Camelin PhD grant is
cofunded by CNRS and Région PACA (Emplois Jeunes Doctorants). 
\end{acknowledgments}


\end{document}